\documentclass[conference]{IEEEtran}
\IEEEoverridecommandlockouts
\usepackage{amsmath,amssymb,amsfonts}
\usepackage{algorithmic}
\usepackage{graphicx}
\usepackage{textcomp}
\usepackage{xcolor}
\usepackage{multirow}
\usepackage{caption}
\usepackage{subcaption}
\usepackage[utf8]{inputenc}

\DeclareUnicodeCharacter{0301}{\'{a}}
\DeclareUnicodeCharacter{0301}{\'{i}}
\begin{document}

\title{Sense and Sensibility: Characterizing Social Media Users Regarding the Use of Controversial Terms for COVID-19\\
}

\author{\IEEEauthorblockN{Hanjia Lyu}
\IEEEauthorblockA{\textit{Goergen Institute for Data Science} \\
\textit{University of Rochester}\\
Rochester, USA \\
hlyu5@ur.rochester.edu}
\and
\IEEEauthorblockN{Long Chen, Jiebo Luo}
\IEEEauthorblockA{\textit{Department of Computer Science} \\
\textit{University of Rochester}\\
Rochester, USA \\
lchen62@u.rochester.edu, jluo@cs.rochester.edu}
\and
\IEEEauthorblockN{Yu Wang}
\IEEEauthorblockA{\textit{Department of Political Science} \\
\textit{University of Rochester}\\
Rochester, USA \\
ywang176@ur.rochester.edu}
}

\maketitle

\begin{abstract}
With the world-wide development of 2019 novel coronavirus, although WHO has officially announced the disease as COVID-19, one controversial term - ``Chinese Virus" is still being used by a great number of people. In the meantime, global online media coverage about COVID-19-related racial attacks increases steadily, most of which are anti-Chinese or anti-Asian. As this pandemic becomes increasingly severe, more people start to talk about it on social media platforms such as Twitter. When they refer to COVID-19, there are mainly two ways: using controversial terms like ``Chinese Virus" or ``Wuhan Virus", or using non-controversial terms like ``Coronavirus". In this study, we attempt to characterize the Twitter users who use controversial terms and those who use non-controversial terms. We use the Tweepy API to retrieve 17 million related tweets and the information of their authors. We find significant differences between these two groups of Twitter users across their demographics, user-level features like the number of followers, political following status, as well as their geo-locations. Moreover, we apply classification models to predict Twitter users who are more likely to use controversial terms. To our best knowledge, this is the first large-scale social media-based study to characterize users with respect to their usage of controversial terms during a major crisis.
\end{abstract}


\section{Introduction}
\begin{figure}[htbp]
\centerline{\includegraphics[scale = 0.18]{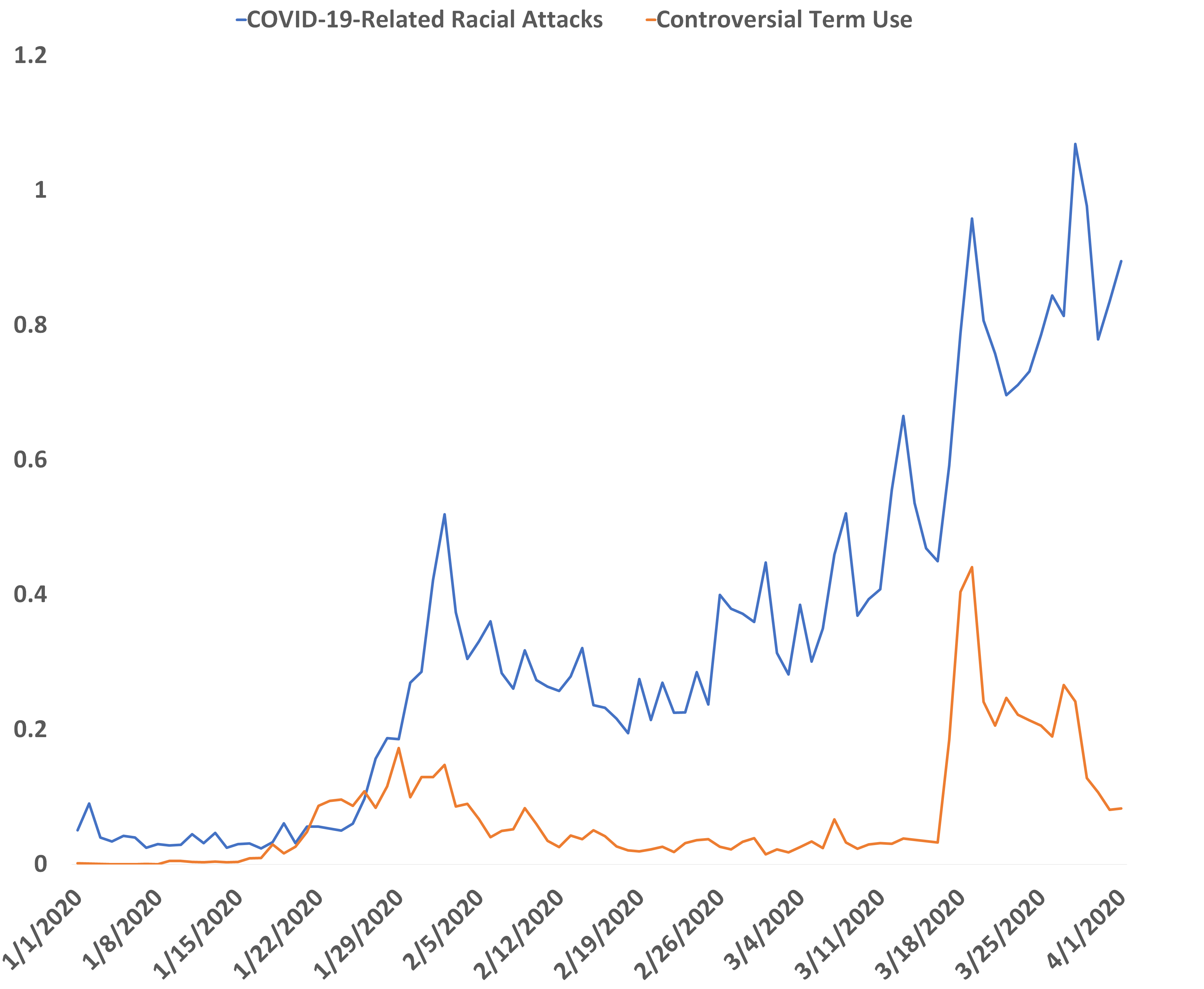}}
\caption{Density of online media coverage with the controversial term and COVID-19 related racial attacks.}
\vspace{-0.2cm}
\label{media_coverage}
\end{figure}

The COVID-19 viral disease was officially declared a pandemic by the World Health Organization (WHO) on March 11. On April 13, WHO reported that 213 countries, areas and territories were impacted by the virus, totaling 1,773,084 confirmed cases and 111,652 confirmed death worldwide.\footnote{https://www.who.int/emergencies/diseases/novel-coronavirus-2019/situation-reports/} This disease has undeniably impacted the daily operations of the society. McKibbin and Fernando provided the estimated overall GDP loss caused by COVID-19 in seven scenarios, with the estimated loss range between 283 billion USD to 9,170 billion USD~\cite{b1}. However, the economy is not the only aspect impacted by COVID-19. When COVID-19 was first spreading in the mainland of China, Lin found that a mutual discrimination was developed within the Asian societies~\cite{b2}. With the world-wide development of COVID-19, the global online media coverage of the term ``Chinese Flu" took off around March 18.\footnote{https://blog.gdeltproject.org/is-it-coronavirus-or-covid-19-or-chinese-flu-the-naming-of-a-pandemic/} Fig.~\ref{media_coverage} shows the timeline of the density of global online media coverage using the term ``Chinese Flu" and the global online media coverage of COVID-19-related racial attacks. Around February 2, there was a peak of the online media coverage of COVID-19-related racial attacks following the peak of online media coverage using ``Chinese Flu". Zheng et al. found that some media coverage of COVID-19 has a negative impact on Chinese travellers' mental health by labeling the outbreak as ``Chinese virus pandemonium"~\cite{b3}. The online media coverage of COVID-19-related racial attacks is still increasing steadily as of April 2020.\footnote{https://blog.gdeltproject.org/online-media-coverage-of-covid-19-related-racial-attacks-increasing-steadily/} Not only the online media coverage but also the usage of the term ``Chinese Virus" or ``Chinese Flu" is trending on social media platforms such as Twitter. On March 16, even the president of United States, Donald Trump posted a tweet calling COVID-19 ``Chinese Virus".\footnote{https://twitter.com/realdonaldtrump/status/1239685852093169664} Despite the defense by President Donald Trump that calling coronavirus the ``Chinese Virus" is not racist\footnote{https://www.cnbc.com/2020/03/18/coronavirus-criticism-trump-defends-saying-chinese-virus.html}, racism and discrimination against Asian-Americans has surged in the US.\footnote{https://www.washingtonpost.com/technology/2020/04/08/coronavirus-spreads-so-does-online-racism-targeting-asians-new-research-shows/}

Matamoros-Fernandez proposed the concept ``platformed racism" as a new form of racism derived from the culture of social media platforms in 2017 and argued that it evoked platforms as amplifiers and manufacturers of racist discourse~\cite{b4}. It is crucial for governments, social media platforms and individuals to understand such phenomena during this pandemic or similar crisis. In addition, the uses of controversial terms associated with COVID-19 could be associated with hate speeches as they are likely xenophobic~\cite{b38}. Hate speeches reflect the expressions of conflicts within and across societies or nationalities~\cite{b39}. On online social media platforms, hate speeches can spread extremely fast, even cross-platform, and can stay online for a rather long time~\cite{b39}. They are also itinerant, meaning that despite forcefully removed by the platforms, they can find expression elsewhere on the Internet and even offline~\cite{b39}. The impact of hate speeches is not to be under-estimated in the sense of both developing racism in the population and damaging the relationship between societies and nations. Therefore, it is crucial to detect and contain the spreading of hate speeches in the early stage to prevent exponential growth of such a mindset among Internet users.

In this study, we attempt to characterize Twitter users who use controversial terms associated with COVID-19 (e.g. ``Chinese Virus") and those who use non-controversial terms (e.g. ``Corona Virus") by analyzing the population bias across demographics, user-level features, political following statuses, and geo-locations. Such findings and insights can be vital for policy-making on the global fight against COVID-19. In addition, we attempt to train classification models for predicting the usage of controversial terms associated with COVID-19, with features crawled and generated from Twitter data. The models can be used in social media platforms to monitor discriminating posts and prevent them from evolving into serious racism-charged hate speeches.

\section{Related Work}

Researches have been conducted to analyze social media users at a general level. Mislove et al. studied the demographics of general Twitter users including their geo-location, gender and race~\cite{b5}. Sloan et al. derived the characteristics of age, occupation and social class from Twitter user meta-data~\cite{b6}. Corbett et al. and Chang et al. also tried to study the demographics of general Facebook users~\cite{b7, b8}. Pennacchiotti and Popescu used a machine learning approach to classify Twitter users according to their user profile, tweeting behavior, content of tweets and network structure~\cite{b9}. With the connection between social media and people's lives getting increasingly closer, studies have been conducted at a more specific level. Gong et al. studied the silent users in social media communities and found that user generated content can be used for profiling silent users~\cite{b10}. Paul et al. were the first to utilize network analysis to characterize the Twitter verified user network~\cite{b11}. Users consciously deleting tweets were observed by Bhattacharya et al.~\cite{b29}. Cavazos-Rehg et al. studied the followers and tweets of a marijuana-focused Twitter handle~\cite{b35}. Moreover, attention has also been paid to users with specific activities. Ribeiro et al. focused on detecting and characterizing hateful users in terms of their activity patterns, word usage and the network structure~\cite{b12}. Olteanu, Weber and Gatica-Perez studied the \#BlackLivesMatter movement and hashtag on Twitter to quantify the population biases across user types and dempographics~\cite{b14}. Badawy et al. analyzed the digital traces of political manipulation related to 2016 Russian interference in terms of Twitter users' geo-location and their political ideology~\cite{b15}. In addition to analyzing individual users, many studies were conducted on communities in social media platforms~\cite{b16,b17,b18}, user behavior~\cite{b20, b21, b22}, and the content that users publish~\cite{b27,b28,b29,b35}.

In our research, we focus on demographics, user-level features, political following status, and the geo-locations of the twitter users using controversial terms and the users using non-controversial terms. The use of a controversial term does not necessarily mean that the user agrees with the term. It also goes the opposite way. For example, consider this tweet:
\begin{center}
    \textit{I don't think we should use the term ``Chinese Virus."}\\
\end{center}

However, we believe the percentage of this kind of tweets is relatively low. We assume that if a user uses the controversial terms, it means this user agrees with the intent of the controversial terms. If a user uses non-controversial terms, then it means this user disagrees with or at least avoids the usage of controversial terms. This assumption allows us to separate the users into two groups according to their choices of the terms: one including those who use controversial terms and the other including those who use non-controversial terms. To the best of our knowledge, this is the first large-scale social media-based study to characterize users with respect to their usage of controversial terms during a major crisis.

\section{Data Collection and Preprocessing}

The related tweets (Twitter posts) were collected using the Tweepy API. We initialized an experimental round to collect posters for 24 hours for both controversial and non-controversial keywords, with “chinese virus”, “china virus” and “wuhan virus” as controversial keywords, and “corona”, “covid19” and “\#Corona” as non-controversial keywords. We then selected the most frequent keywords and hashtags in the controversial dataset and the non-controversial dataset for a refined keyword list. As a result, the controversial keywords consist of “chinese virus” and “\#ChineseVirus” and non-controversial keywords included “corona”, “covid-19”, “covid19”, “coronavirus”, “\#Corona”, “\#Covid\_19” and “\#coronavirus”.\footnote{In Tweepy query, capitalization of non-hashtag keywords does not matter.} The keywords were then used to crawl tweets to construct a dataset of controversial tweets (CD) and a dataset of non-controversial ones (ND) simultaneously for a four-day period from March 23-26, 2020. In the end, 1,125,285 tweets were collected for CD and 16,320,176 for ND. We created four pairs of CD-ND datasets as follows.

\subsection{Baseline Datasets}

We built the baseline datasets with basic attributes that can be used in all subsequent subsets. In the Baseline Datasets, 7 user-level features were either collected or computed, including followers\_count, friends\_count, statuses\_count, favorites\_count, listed\_count, account\_length (the number of months since the account was created) and verified status (verified users are the accounts that are independently authenticated by the platform\footnote{https://developer.twitter.com/en/docs/tweets/data-dictionary/overview/user-object} and are considered influential). Next, entries with missing values were removed. In the end, 7 features were computed for 1,125,176 tweets in CD and 1,599,013 tweets in ND. CD and ND were quite balanced with random sampling for convenience in classification process. In addition, since our analysis were performed with proportion tests, the randomly sampling process still maintained representation for the entire dataset.

Since our paper focuses on user-level features, and most of which remained unchanged for a user throughout the 4-day data collection period, we removed duplicate users in CD and ND, respectively, to reduce duplicate entries in our datasets. However, we did not remove duplicate users that appeared in both CD and ND, as the number of such users were unsubstantial (8.19\% of the total users) and such ``swing users" (users who used both controversial and non-controversial terms) were a potential type of users that we definitely should not exclude. In the end, there are 593,233 distinct users in CD and 490,168 distinct users in ND.

\begin{table}[htbp]
\caption{Composition of Profile Images.}
\vspace{-0.2cm}
\begin{center}
\begin{tabular}{c c c}
\hline
 & Controversial & Non-Controversial\\
 \hline
One Intelligible Face & 47,011 & 109,718\\
Multiple Faces & 5,596 & 11,393\\
Zero Intelligible Face & 54,539 & 96,218\\
Invalid URL & 264,894 & 187,379\\
\hline
Total & 372,040 & 404,708\\
\hline
\end{tabular}
\label{demo}
\end{center}
\vspace{-0.25cm}
\end{table}

\subsection{Demographic Datasets}

We intend to investigate user-level features of the Twitter users with the demographic datasets, which were built upon the baseline datasets. Since Twitter does not provide sufficient demographic information in the crawled data, we applied Face++ API\footnote{https://www.faceplusplus.com} to obtain inferred age and gender information by analyzing users' profile images. Profile images with multiple faces were excluded. We also found that some profile images were not real-person images and many URLs were invalid. Such data were considered noise and removed from the demographic datasets. Table~\ref{demo} shows the counts of images with one intelligible face, multiple faces, zero intelligible face and invalid URL. Images with only one intelligible face were retained to form the demographic datasets. 

Politically related attributes by tagging users were also added if they follow the Twitter accounts of top political leaders who are or were pursuing nomination for the 2020 presidential general election. Five Democratic presidential candidates (Joe Biden, Michael Bloomberg, Bernie Sanders, Elizabeth Warren and Pete Buttigieg) and the incumbent Republican President (Donald Trump) were included in the analysis. We crawled Twitter followers' IDs for the political figures to determine the respective political following statuses.\footnote{Due to limitation of Twitter API, only about half of Donald Trump's follower ID was crawled.}

In the end, the Demographic Datasets consist of 15 features (7 features from the Baseline Datasets and the 8 aforementioned new features), with 47,011 distinct users in CD and 109,718 distinct users in ND.

\subsection{Geo-location Datasets}

\begin{figure}[!t]

\centerline{\includegraphics[scale = 0.2]{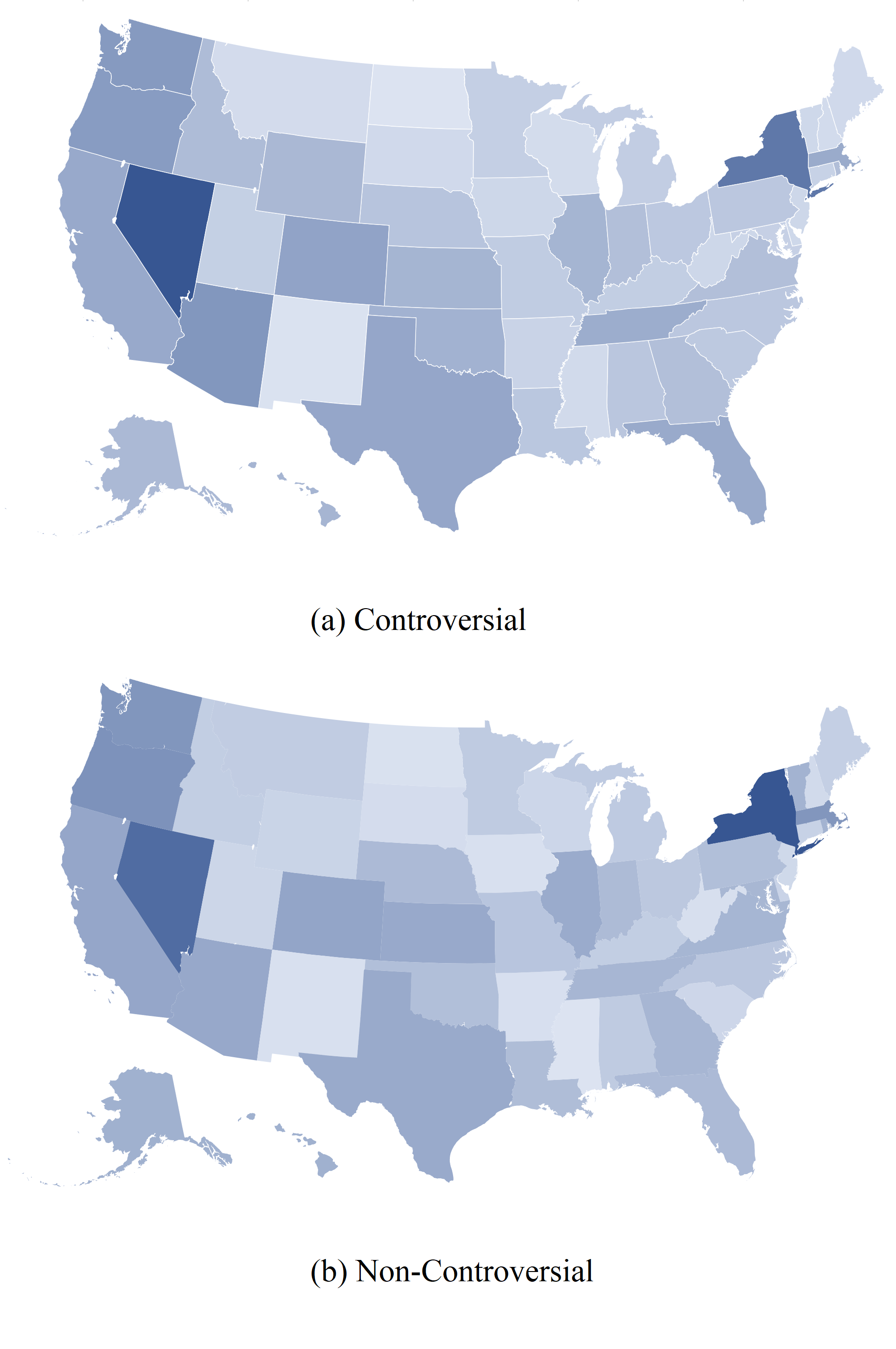}}
\caption{Distribution of a) controversial and b) non-controversial tweets in the US, by state, and normalized by the state population. No significant differences can be observed. It is interesting to note that New York and Nevada have the highest number of COVID-19 related tweets per capita. } 
\label{fig:state_level_distri}
\vspace{-0.25cm}
\end{figure}

We also intend to investigate the potential impact of the type of communities where users live in on their uses of controversial terms associated with COVID-19. Therefore, Geo-location Datasets were built upon the Baseline Datasets, in a similar fashion as the Demographic Datasets.

Observing that only a very limited number of tweets contain self-reported locations (1.2\% of crawled data),  we instead use the user profile location as the source of geo-location, which has a substantially higher percentage of entries in the crawled datasets (16.2\% of crawled data). We collected posts with user profile location entries and then removed entries that are clearly  noise (e.g. ``Freedom land", ``Moon" and ``Mars"), non-US locations and unspecific locations (ones that only report country or state). In the end, there are 14,817 users for CD and 41,118 users for ND in the Geo-location Datasets.

At the state level, we observe little difference in the distribution of tweets between CD and ND, as shown in Fig.~\ref{fig:state_level_distri}. Therefore, a more detailed analysis of geo-location information is required. Detailed information about locations was collected with python package uszipcode. Geo-location data in the datasets were processed to find the exact location at zip-code level. Based on population density of a zip code area, we then classified locations into urban (3,000+ persons per square mile), suburban (1,000 – 3,000 persons per square mile) or rural (less than 1,000 persons per square mile).\footnote{https://greatdata.com/product/urban-vs-rural}

\subsection{The Aggregate Datasets}

Datasets with both demographic and geo-location features were created. These datasets contain complete attributes that were analyzed in our study, while trading off with the relatively small size, with 5,772 for CD and 12,403 for ND. These datasets can be used in a classification model to compare feature importance among all attributes.

\section{Characterizing Users Using Different Terms}

We perform statistical analysis with the generated datasets in order to investigate and compare patterns of features in both CD and ND for demographic, user-level, political and location-related attributes.

\begin{figure}[htbp]
\centerline{\includegraphics[scale = 0.7]{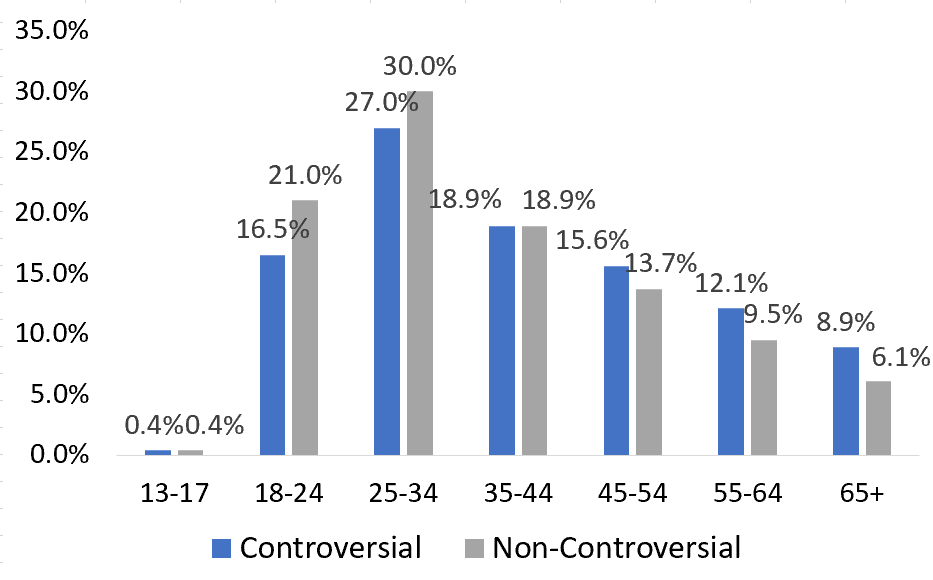}}
\caption{Age Distribution among users of controversial terms and users of non-controversial terms.}
\label{age}
\vspace{-0.2cm}
\end{figure}

\subsection{Demographic Analysis}

\subsubsection{Age}
Fig.~\ref{age} is the distribution of age. We separate the age range into seven bins similar to most social media analytic tools. In both CD and ND, the 25-34 bin comprises the biggest part, which is consistent with the age distribution of general Twitter users.\footnote{https://www.statista.com/statistics/283119/age-distribution-of-global-twitter-users/} After performing the goodness-of-fit test, we find that the age distributions in these two groups are statistically different ($p<0.0001$). The Twitter users in ND tend to be younger. More than half of the non-controversial group are the people under 35 years old. In ND, there are 21.0\% users in the 18-24 bin while that proportion is only 16.5\% in CD. Users that are older than 45 years old are more likely to use controversial terms.

\begin{figure*}[htbp]
\centerline{\includegraphics[scale = 0.088]{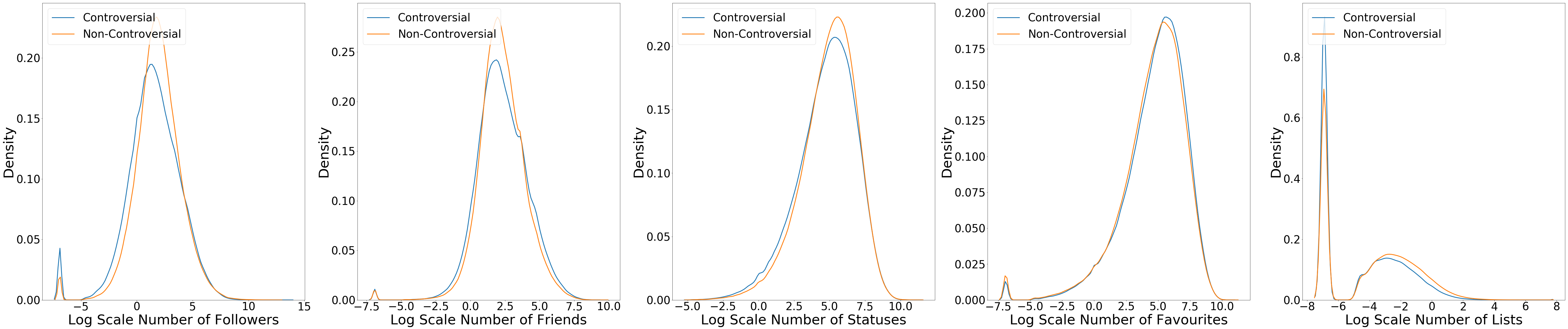}}
\caption{Density plots (log-scale) of the normalized numbers of followers, friends, statuses, favourites, and listed.}
\label{user_level_plot}
\end{figure*}

\subsubsection{Gender}
Table~\ref{gender} shows the gender distribution of each group. In both groups, there are more male. There are 61.0\% male users in CD and 56.2\% male users in ND. As of January 2020, 62\% Twitter users are male, and 38\% are female.\footnote{https://www.statista.com/statistics/828092/distribution-of-users-on-twitter-worldwide-gender/} This observation suggests that the gender distribution of CD is not different from the overall Twitter users. Furthermore, We perform the proportion z-test and there is sufficient evidence to conclude that gender distributions in these two groups are statistically different ($p<0.0001$). There are relatively more female users in ND which account for 46.1\%, however there are only 38.0\% female users in CD.

\begin{table}[htbp]
\caption{Gender Distribution.}
\vspace{-0.1cm}
\begin{center}
\begin{tabular}{c c c}
\hline
Gender & Controversial & Non-Controversial \\
 \hline
Male & 61.0\% &  56.2\% \\
Female & 39.0\% & 43.8\% \\
\hline
\end{tabular}
\label{gender}
\end{center}
\vspace{-0.2cm}
\end{table}

\subsection{User-level Features}
In this subsection, we attempt to find insights into 7 user-level features including the number of followers, friends, statuses, favourites, listed, and the number of months since the user account was created, and the verified status.  The number of statuses retrieved using Tweepy is the count of the tweets (including retweets) posted by the user. The number of listed is the number of public lists that this user is a member of. 

To better analyze the number of followers, friends, statuses, favorites and listed, we normalize them by the number of months since the user's account was created. Given the domain range of these five attributes is large and to better observe the distribution, we first add 0.001 to all these values to avoid zero probability and take the logarithm of them. Fig.~\ref{user_level_plot} shows the density plots (in log scale) of the normalized numbers of followers, friend, statuses, favourites, and listed. Since the distribution is not close to a normal distribution, we apply the Mann-Whitney rank test on these five attributes. There is strong evidence ($p<0.0001$) in all these five attributes to conclude that the respective medians of these features of CD are not equal to the ones of ND. Table \ref{median} shows the medians of the five normalized features in each group. Users using non-controversial terms tend to have a larger social capital which means they have relatively more followers, friends and post more tweets. This suggests that users in ND normally have a larger size of audience and are more active in posting tweets. One hypothesis for this is that users with a larger audience and more experienced with using Twitter are more cautious when posting in Twitter, which means they pay more attention to the choice of words. Twitter users are found to be more cautious in publishing and republishing tweets, and also more cautious in sharing among friends\cite{b44}. Although the medians of favourites and listed memberships of ND are also higher than those of CD, the differences are not large.

\begin{table}[htbp]
\caption{Medians of numbers of followers, friends, statuses, favourites, and listed.}
\vspace{-0.1cm}
\begin{center}
\begin{tabular}{c c c}
\hline
Features & Controversial & Non-Controversial \\
 \hline
\# of Followers & 227  &  360\\
\# of Friends  & 413 &  494\\
\# of Statuses & 6617  & 9241\\
\# of Favourites & 7681  & 7860.5 \\
\# of Listed & 1  & 2\\
\hline
\end{tabular}
\label{median}
\end{center}
\vspace{-0.2cm}
\end{table}

\begin{table}[htbp]
\caption{Distribution of verified users.}
\vspace{-0.1cm}
\begin{center}
\begin{tabular}{c c c}
\hline
User Type & Controversial & Non-Controversial \\
 \hline
Verified Users & 0.6\% &  2.0\% \\
Non-Verified Users & 99.4\% & 98.0\% \\
\hline
\end{tabular}
\label{verified}
\end{center}
\end{table}

\begin{figure}[htbp]
\centerline{\includegraphics[scale = 0.23]{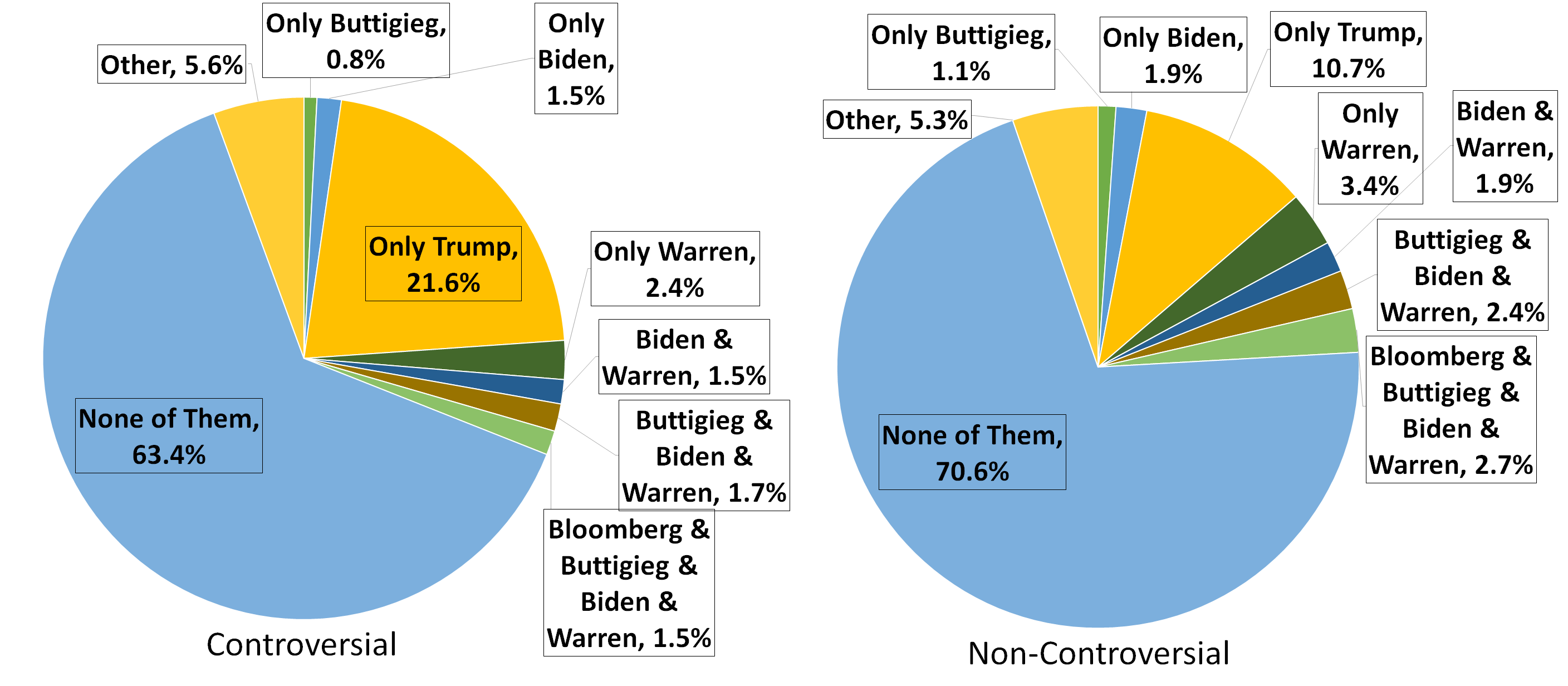}}
\caption{Proportion of Political Following Status.}
\label{political}
\vspace{-0.2cm}
\end{figure}

In addition, the users using controversial terms tend to have newer accounts. By applying the Mann-Whitney rank test, the number of months since account was created by users in CD is statistically smaller ($p<0.0001$). The median number of months of CD is 63 and that of ND is 74. The difference is almost one year, which is significant. This observation is similar to the findings that hateful users have newer accounts\cite{b12}. Since using controversial terms does not necessarily correspond to delivering hateful speech, we hypothesize that newer accounts could indicate less experience using Twitter and thus being less cautious when posting tweets.



According to the latest statistics report, there were 0.05\% verified users overall.\footnote{https://www.digitaltrends.com/social-media/journalists-celebrities-athletes-make-up-most-of-twitters-verified-list/} Although there are more non-verified users in each group, the proportions of the verified users in CD and ND are both higher than that of the overall Twitter population. After performing a proportion z-test, we find that the proportion of verified users in ND is greater than that in CD ($p<0.0001$). Table \ref{verified} is the distribution of verified users. This indicates that verified users are more likely to use non-controversial terms. Existing work shows the tweets posted by verified users are more likely to be perceived as trustworthy~\cite{b41} which leads to our hypothesis that verified users are more cautious publishing tweets since their tweets have more credibility partially due to their conscious efforts.

\subsection{Political Following Status}

To model the political following status, we record if the user is a follower of Joe Biden, Bernie Sanders, Pete Buttigieg, Michael Bloomberg, Elizabeth Warren, or Donald Trump. There are 64 different combinations of following behaviors among all the 1,083,401 Twitter users in our dataset. Fig.~\ref{political} is the pie chart of political following status, where we only show the proportions of the combinations greater than 1.0\%. By performing the goodness-of-fit test, we find statistical difference with respect to the proportions ($p<0.0001$). In both groups, most of users do not  follow any of these people. This group constitutes 70.6\% in ND and 63.4\% in CD, respectively. The second biggest group of users in both CD and ND corresponds to users who only follow Donald Trump. The proportion of users in CD only following Donald Trump is 21.6\% which is higher than that in ND. This indicates that users only following President Trump are more likely to use controversial terms. This observation resonates with the findings by Gupta, Joshi, and Kumaraguru~\cite{b16} that during a crisis, the leader represents the topics and opinions of the community. Another noticeable finding is the proportions of users that follow the members of the Democratic Party in ND are all higher than those in CD. This suggests that users following the members of the Democratic Party are more likely to use non-controversial terms. 

\subsection{Geo-location}

Geo-location is potentially another important factor in the use of controversial terms. As shown in Table~\ref{tab:geo}, the proportion of controversial tweets by users from rural or suburban regions is significantly higher than that of non-controversial tweets ($p<0.0001$). The result suggests that Twitter users in rural or suburban regions are more likely to use controversial terms when discussing COVID-19, where as Twitter users in urban areas tend to use non-controversial ones in the same discussion. There could be a number of reasons behind this.

One explanation is the difference in political views between metropolitan and non-metropolitan regions. Especially after the 2016 presidential election, political polarization in a geographical sense has been increasing dramatically. Scala et al. pointed out that in the 2016 election, most urban regions supported Democrats, whereas most rural regions supported Republicans~\cite{b42}. The study also showed that variations in voting patterns and political attitudes exist along a continuum, meaning that the Democratic-Republican supporter ratio gradually changes from metropolitan to rural areas~\cite{b42}. With our finding in the previous section that Republican-politician Twitter followers have a higher tendency to use controversial terms, the difference in usage geographically could be explained. However, Scala et al. also pointed out that suburban areas were still heavily Democratic in the 2016 election~\cite{b42}. Thus we cannot fully explain the overwhelming controversial usage in suburban regions from a pure political perspective.

Another explanation is that more urbanized regions are more affected by COVID-19. Rocklöv et al. discovered in a recent research that the contract rate of the virus is proportional to population density, resulting in significantly higher basic reproduction number\footnote{Basic reproduction number (R0) is the expected number of cases directly generated by one case in a population where all individuals are susceptible to infection.} and thus more infections in urban regions than in rural regions~\cite{b43}. Therefore, higher reporting of infections cases can be found in urban regions, contributing to the higher use of non-controversial terms associated with COVID-19. On the contrary, rural regions have relatively lower percentage of infected population, thus discussions about the nature and the origin of COVID-19 could be more prevalent, and during such discussions, relatively more users chose to use controversial terms. This could partially explain the higher use of controversial terms in suburban regions.

\begin{table}[htbp]
\caption{Tweet percentages in urban, suburban and rural areas.}
\begin{center}
\begin{tabular}{c c c}
\hline
 & Controversial & Non-Controversial\\
 \hline
Urban & 56.14\% & 62.48\% \\
Suburban & 17.48\% & 15.58\% \\
Rural & 26.38\% & 21.94\% \\
\hline
\end{tabular}
\label{tab:geo}
\end{center}
\vspace{-0.2cm}
\end{table}

\section{Modeling Users of controversial terms}
One goal of this paper is to predict Twitter users who employ controversial terms in the discussion of COVID-19 using demographic, political and post-level features that we crawled and generated in the previous sections. Therefore, various regression and classification models were applied. In total, six models were deployed: Logistic Regression (Logit), Random Forest (RF), Support Vector Machine (SVM), Stochastic Gradient Descent Classification (SGD) and Multi-layer Perceptron (MLP) from the Python sklearn package, and XGBoost Classification (XGB) from the Python xgboost package. For SVM, a linear kernel was used. For SGD, the Huber loss was used to optimize the model. For MLP, three hidden layers were used with sizes of 150, 100 and 50 nodes respectively, the Rectified Linear Unit (ReLU) was used as activation for all layers, and the Adam Optimizer was used for optimization. The models were applied to three datasets as described below. All datasets were split into training and testing sets with a ratio of 90:10. Since hyperparameter tuning was done in an empirical fashion, no development set was used. The data split was unchanged for models in each experiment. In addition, non-binary data were normalized using min-max normalization.

\begin{table*}[h]
    \centering
    \begin{subtable}[h]{0.35\textwidth}
        \centering
            \begin{tabular}{c|ccc}
            \hline
            \multirow{2}{*}{\textbf{Metric}} & \multicolumn{3}{c}{\textbf{Baseline}}\\
            \cline{2-4}
             
            & \textbf{Logit} & \textbf{XGB} & \textbf{RF}\\
             \hline
            \textbf{Accuracy} & 0.625 & 0.628 & 0.645\\
            \textbf{Precision} & 0.604 & 0.594 & 0.606\\
            \textbf{Recall} & 0.268 & 0.315 & 0.396\\
            \textbf{F1} & 0.371 & 0.411 & 0.479\\
            \textbf{AUC-ROC} & 0.618 & 0.647 & \textbf{0.669}\\
            
            \hline
            \end{tabular}
       \caption{Baseline (N\textsubscript{CD}=593,233 , N\textsubscript{ND}=490,168)}
       \label{tab:classifier_baseline}
    \end{subtable}
    \hspace{0.01em}
    \begin{subtable}[h]{0.4\textwidth}
        \centering
            \begin{tabular}{c|cccccc}
            \hline
            \multirow{2}{*}{\textbf{Metric}} & \multicolumn{6}{c}{\textbf{Demographics}} \\
            \cline{2-7}
             
            & \textbf{Logit} & \textbf{XGB} & \textbf{RF} & \textbf{SVM} & \textbf{MLP} & \textbf{SGD}\\
             \hline
            \textbf{Accuracy} & 0.584 & 0.691 & 0.743 & 0.510 & 0.680 & 0.627\\
            \textbf{Precision} & 0.604 & 0.698 & 0.735 & 0.474 & 0.663 & 0.665\\
            \textbf{Recall}  & 0.296 & 0.589 & 0.698 & 0.544 & 0.631 & 0.390\\
            \textbf{F1} & 0.397 & 0.639 & 0.716 & 0.507 & 0.647 & 0.491\\
            \textbf{AUC-ROC} & 0.633 & 0.783 & \textbf{0.833} & 0.699 & 0.714 & 0.783\\
            
            \hline
            \end{tabular}
        \caption{Demographics (N\textsubscript{CD}=47,011, N\textsubscript{ND}=109,718)}
        \label{tab:classifier_demo}
     \end{subtable}
     
     \par\bigskip
     
    \begin{subtable}[h]{0.5\textwidth}
        \centering
            \begin{tabular}{c|cccccc}
            \hline
            \multirow{2}{*}{\textbf{Metric}} & \multicolumn{6}{c}{\textbf{Geo-location}}\\
            \cline{2-7}
             
            & \textbf{Logit} & \textbf{XGB} & \textbf{RF} & \textbf{SVM} & \textbf{MLP} & \textbf{SGD}\\
             \hline
            \textbf{Accuracy} & 0.611 & 0.651 & 0.811 & 0.543 & 0.615 & 0.658\\
            \textbf{Precision} & 0.695 & 0.699 & 0.757 & 0.508 & 0.713 & 0.702\\
            \textbf{Recall} & 0.305 & 0.436 & 0.875 & 0.499 & 0.504 & 0.456\\
            \textbf{F1} & 0.422 & 0.537 & 0.812 & 0.504 & 0.410 & 0.553\\
            \textbf{AUC-ROC} & 0.652 & 0.696 & \textbf{0.874} & 0.696 & 0.636 & 0.713\\
            
            \hline
            \end{tabular}
       \caption{Geo-location (N\textsubscript{CD}=14,817, N\textsubscript{ND}=41,118)}
       \label{tab:classifier_geo}
    \end{subtable}
    \hspace{0.01em}
    \begin{subtable}[h]{0.4\textwidth}
        \centering
            \begin{tabular}{c|cccccc}
            \hline
            \multirow{2}{*}{\textbf{Metric}} & \multicolumn{6}{c}{\textbf{Demographics + Geo}}\\
            \cline{2-7}
             
            & \textbf{Logit} & \textbf{XGB} & \textbf{RF} & \textbf{SVM} & \textbf{MLP} & \textbf{SGD}\\
             \hline
            \textbf{Accuracy} & 0.628 & 0.592 & 0.644 & 0.591 & 0.685 & 0.658\\
            \textbf{Precision} & 0.425 & 0.402 & 0.439 & 0.328 & 0.495 & 0.476\\
            \textbf{Recall} & 0.524 & 0.622 & 0.497 & 0.293 & 0.264 & 0.241\\
            \textbf{F1} & 0.469 & 0.489 & 0.467 & 0.310 & 0.344 & 0.319\\
            \textbf{AUC-ROC} & \textbf{0.624} & 0.545 & 0.604 & 0.510 & 0.571 & 0.560\\
            
            \hline
            \end{tabular}
        \caption{Aggregate (N\textsubscript{CD}=5,772, N\textsubscript{ND}=12,403)}
        \label{tab:classifier_aggregate}
     \end{subtable}
     \caption{Classification metrics for the four pairs of datasets. Best AUC-ROC scores are highlighted in each dataset pair.}
     \label{tab:classifier_result}
\end{table*}

\subsection{Classification on Baseline Datasets}

7 attributes were collected in the baseline datasets for CD and ND. For classification, we removed the favorites count and listed count since they were found to be trivial in our preliminary analysis. In the end, 5 features (follower count, friend count, status count, account length and verified status) were included in the training dataset.

\subsection{Classification on Demographic Datasets}

With the demographic datasets, user-level (political following, age and gender) features were included. We converted age into dummy variables, with cutoffs as reported in Section IV. Gender was also converted into dummy variables \textit{female} and \textit{male}. In the end, 18 features were prepared as inputs for the machine learning models. In total, 21 variables were created.

\subsection{Classification on Geo-location Datasets}

In addition to the 5 features in the baseline datasets, geolocation classification (urban, suburban and rural) was included in geo-location classification, resulting in 8 features in the datasets for geo-location classification. 

\subsection{Classification on Aggregate Datasets}

For the aggregate datasets, all of the 24 aforementioned features in the previous three datasets were included. We aimed to see comparisons of importance and impact of features on the use of controversial terms.
\begin{figure*}[htbp]
\centering
\includegraphics[width=0.88\linewidth]{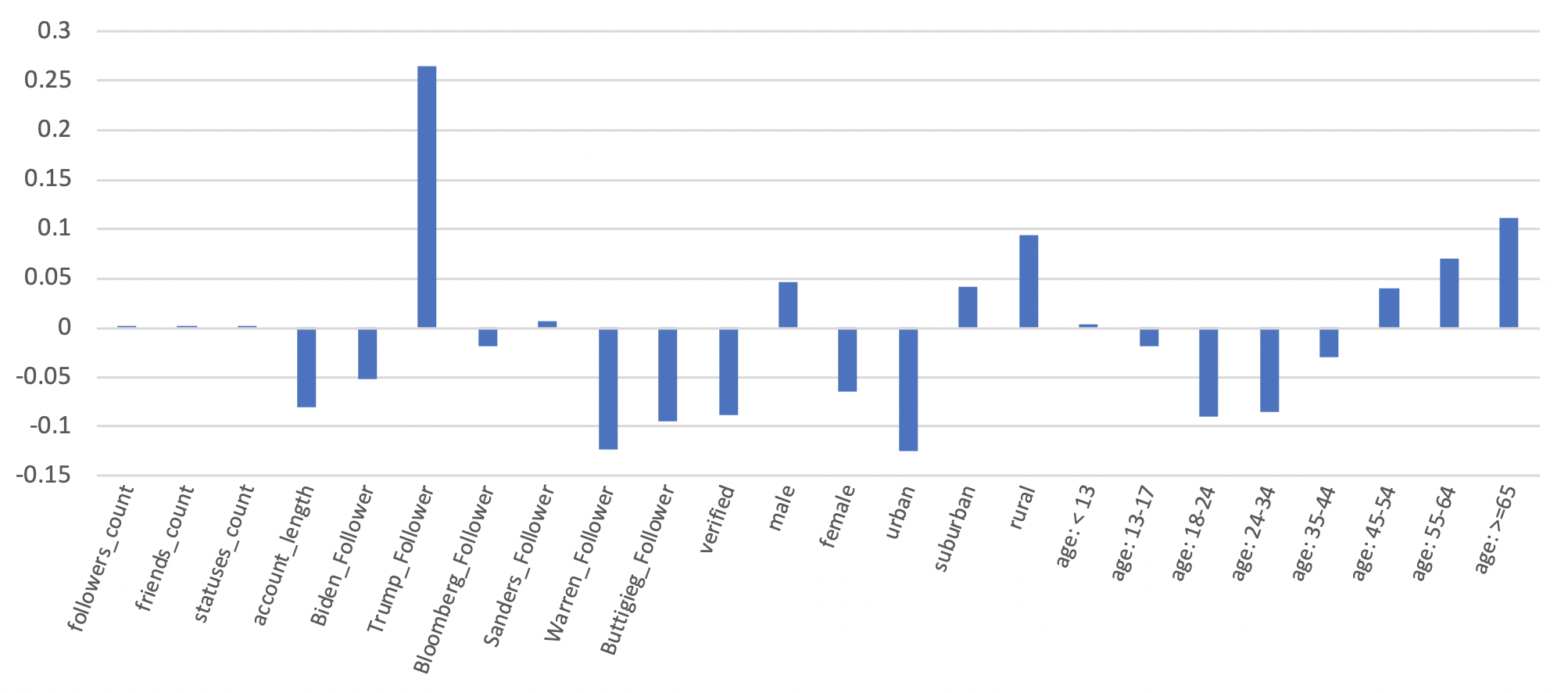}
\vspace{-0.2cm}
\caption{Logistic regression coefficients for the aggregate dataset. Users who follows Trump, users in rural and suburban regions, male users and users aged over 45 years old tend to use controversial terms, whereas users who follow most top Demogratic presidental candidates, users in urban regions, female users and users aged between 18 to 44 tend to use non-controversial terms. No significant association between account-level attributes (follower, friend and status count) and the use of controversial terms were found.}
\label{fig:model_coef}
\vspace{-0.2cm}
\end{figure*}
\subsection{Results and Evaluations}

The classification metrics of the datasets are shown in Table \ref{tab:classifier_result}. In general, Random Forest classifiers resulted in the best results in terms of the AUC-ROC measures. The best AUC-ROC score achieved in baseline dataset classification is 0.669, that in demographic dataset classification is 0.833, that in geo-location dataset classification is 0.874 and that in aggregate dataset classification is 0.624. The AUC-ROC scores in demographic and geo-location classification are rather high, showing strong signals in demographic and geo-location related features.
The AUC-ROC score in the baseline datasets is understandably low since only 5 features were included. Model performance in the aggregate datasets is also relatively low. A possible explanation is the small training data size compared to the demographic and geo-location datasets.

The impact of features on predicting the use of controversial terms can be observed using logistic regression coefficients, as shown in Fig.~\ref{fig:model_coef}. The most significant contribution to predicting the use of controversial terms is Trump\_Follower, while the followings of democratic party politicians contributes negatively. Male users tend to use controversial terms, while female users tend to use non-controversial terms indicated by the higher absolute coefficient. Urban users tend to use non-controversial terms, which suburban and rural users have higher uses of controversial ones. Verified users tend to use non-controversial terms. Accounts with longer history tend to use non-controversial terms. Users under 45 years old tend to use non-controversial terms, especially for those between 18 to 35, while users older than 45 years old tend to use controversial terms. In addition, user-level attributes (following count, friends count and statuses count) have very little impact in the model. These findings are in principle comparable to the analysis in previous sections.

\section{Conclusions and Future Work}

We have analyzed 1,083,401 Twitter users in terms of their usage of terms referring to COVID-19. We find significant differences between the users using controversial terms and users using non-controversial terms across demographics, user-level features, political following status, and geo-location. Young people tend to use non-controversial terms to refer to  COVID-19. Although in both CD and ND groups, male users constitute a higher proportion, the proportion of female users in the ND group is higher than that in the CD group. In terms of user-level features, users in the ND group have a larger social capital which means they have more followers, friends, and statuses. In addition, the proportion of non-verified users is higher in both CD and ND groups, while the proportion of verified users in the ND group is higher than that of the CD group. There are more users following Donald Trump in the CD group than in the ND groups. The proportion of users in the ND group following the members of the Democratic Party is higher. There is no sufficient evidence to conclude that there is difference in  terms of which state the user lives in. However, we find users living in rural or suburban areas are more likely to use the controversial terms than users living in urban areas. Furthermore, we apply several classification algorithms to predict the users with a higher probability of using the controversial terms. Random Forest produces the highest AUC-ROC score of 0.874 in geo-location classification and 0.833 in demographics classification.

Since high accuracy is achieved in both demographics and geo-location classification, future studies may collect larger datasets for aggregate classification for better model performance. In addition, this research mainly focuses on user-level attributes in understanding the characteristics of users who use controversial terms associated with COVID-19. Analysis from other perspectives, such as textual mining for the Twitter posts associated with controversial term uses, can be performed to gain more insights of the use of controversial terms and better understanding of those who use them on social media.

\end{document}